\documentclass{aa}

\usepackage[tbtags,fleqn]{amsmath}
\usepackage{amsfonts}
\usepackage{graphics}
\usepackage{natbib}

\newcommand{\e}{\mathrm e}
\newcommand{\diff}{\mathrm d}

\newcommand{\Lp}{\mathcal L}

\newcommand{\mincir}{\raise
  -2.truept\hbox{\rlap{\hbox{$\sim$}}\raise5.truept \hbox{$<$}\ }}
\newcommand{\magcir}{\raise
  -2.truept\hbox{\rlap{\hbox{$\sim$}}\raise5.truept \hbox{$>$}\ }}

\DeclareMathOperator{\Cov}{Cov}

\begin{document}

\defcitealias{2001A&A...373..359L}{Paper~I}
\defcitealias{2002A&A...392.1153L}{Paper~II}
\allowdisplaybreaks[2]

\title{Smooth maps from clumpy data: generalizations}
\author{Marco Lombardi and Peter Schneider}
\offprints{M. Lombardi}
\mail{lombardi@astro.uni-bonn.de}
\institute{%
  Institut f\"ur Astrophysik und Extraterrestrische Forschung,
  Universit\"at Bonn, Auf dem H\"ugel 71, D-53121 Bonn, Germany}
\date{Received ***date***; Accepted ***date***}
\abstract{%
  In a series of papers \citep{2001A&A...373..359L,
    2002A&A...392.1153L} we studied in detail the statistical
  properties of an interpolation technique widely used in astronomy.
  In particular, we considered the average interpolated map and its
  covariance under the hypotheses that the map is obtained by
  smoothing unbiased measurements of an unknown field, and that the
  measurements are uniformly distributed on the sky.  In this paper we
  generalize the results obtained to the case of observations carried
  out only on a finite field and distributed on the field with a
  non-uniform density.  These generalizations, which are required in
  many astronomically relevant cases, still allow an exact, analytical
  solution of the problem.  We also consider a number of properties of
  the interpolated map, and provide asymptotic expressions for the
  average map and the two-point correlation function which are valid
  at high densities.  }

\maketitle

%

\section{Introduction}
\label{sec:introduction}

Interpolation techniques play a central role in many physical
sciences.  In fact, experimental data can often only be obtained at
discrete points, while quantitative, global analyses can normally be
performed only on a field.  A classical example of such a situation is
given by meteorological data, such as temperature, pressure, or
humidity: these data are collected by a large number of ground-based
weather stations, and then need to be interpolated in order to obtain
a continuous field.  

The situation is, apparently, very similar in Astronomy.  Indeed, many
astronomical observations are carried out ``discretely,'' i.e.\ data
are available only on some locations of the sky (typically
corresponding to some astronomically significant objects, such as
stars, galaxies, quasars).  If there is some reason to think that the
data represent discrete measurements of a continuous field, then the
observer will want to interpolate the data in order to obtain a smooth
map of the quantity being investigated.

In reality, astronomical observations have a characteristic that make
them quite peculiar with respect to other physical experiments: in
most cases, it is not possible to choose where to perform the
measurements.  A meteorologist, for example, can always decide to put
his weather station in a particular location, or to have weather
stations regularly spaced; this, clearly, is impossible for an
astronomer.  As a result, it is sensible to perform a statistical
analysis of interpolation techniques by considering the measurement
locations as random variables, i.e. by performing an ensemble average
on the positions of the astronomical objects used for the analysis
\citetext{this technique has been already used by several authors;
  see, e.g., \citealp{1998A&A...335....1L};
  \citealp{2000MNRAS.313..524W}}.

In a series of previous papers (\citealp{2001A&A...373..359L},
hereafter Paper~I, and \citealp{2002A&A...392.1153L}, hereafter
Paper~II), we analyzed the statistical properties of a widely used
interpolation technique.  In particular, we considered a set of
measurements $\{ \hat f_n \}$ performed at locations $\{ \vec\theta_n
\}$.  The measurements were taken to be unbiased estimates of a field
$f(\vec\theta)$ at the relative positions, i.e.\ 
\begin{equation}
  \label{eq:1}
  \langle \hat f_n \rangle = f(\vec\theta_n) \; ,
\end{equation}
where the brackets $\langle \cdot \rangle$ denote the expectation value of the
enclosed quantity.  The discrete measurements $\{ \hat f_n \}$ were then
interpolated using the following technique.  First, we introduced a
positive function $w\bigl( \vec \phi \bigr)$, which describes the
``influence'' of measurements performed at $\vec\theta' = \vec\theta + \vec\phi$ on
the interpolated field $\tilde f(\vec\theta)$.  This field was defined as
\begin{equation}
  \label{eq:2}
  \tilde f(\vec\theta) \equiv \dfrac{\sum_{n=1}^N \hat f_n
  w(\vec\theta - \vec\theta_n)}{\sum_{n=1}^N w(\vec\theta -
  \vec\theta_n)} \; , 
\end{equation}
where $N$ is the total number of observations.  Equation~\eqref{eq:2},
indeed, is a standard interpolation technique \citep[see,
e.g.][]{Cressie, Lam} called ``moving weights,'' ``moving average,''
or ``distance weighted average'' [this last name is due to the fact
that normally the weight function $w$ used in Eq.~\eqref{eq:2} depends
only on the distance $\| \vec\theta - \vec\theta_n \|$).  Although other
techniques are clearly available \citep[see,
e.g.][]{1996MNRAS.279..693B, 2000A&A...363L..29S,
  2002A&A...395..733L}, this is probably the interpolation method most
often used in Astronomy.

In this paper we study the expectation value and the two-point
correlation of the smoothed map $\tilde f(\vec\theta)$ under the hypotheses
that:
\begin{enumerate}
\item The measurements $\{ \hat f_n \}$ are unbiased estimates of the
  field $f$ [cf.\ Eq.~\eqref{eq:1}] with errors.  The errors $\epsilon_n =
  \hat f_n - f(\vec\theta_n)$ are taken to be independent random variables
  with vanishing mean (this is clearly equivalent to the unbiasness of
  $\{ \tilde f_n \}$):
  \begin{align}
    \label{eq:3}
    \bigl\langle \epsilon_n \bigr\rangle & {} = 0 \; , &
    \bigl\langle \epsilon_n \epsilon_m \bigr\rangle & {} = \delta_{nm} \sigma^2(\vec\theta_n) \; .
  \end{align}
\item The measurement locations are independent random variables
  distributed according to a know density field $\rho(\vec\theta)$ inside a
  given observation area $\Omega$ (i.e., are a non-homogeneous Poisson
  process).
\end{enumerate}
Hence, this work generalizes the results obtained in Paper~I and II in
three directions: (\textit{i\/}) a \textit{non-constant variance\/}
$\sigma^2(\vec\theta)$ is considered; (\textit{ii\/}) the density of measurement
locations $\rho$ can vary on the field; (\textit{iii\/}) no restriction
is put on the size of the observation area $\Omega$, which can be finite.
Surprisingly, although these generalizations significantly widens the
applicability of our results, they do not make our method more
complicated at all.  Indeed, as we will see below, the problem seems
to find a very natural description in the more general framework
used here.

It should be stressed that the generalizations carried out here are
very important in the astrophysical context.  Indeed, astronomical
data will normally be available only on a limited area of the sky, and
so boundary effects have to be taken into account.  Moreover, often
the measurements will not be uniformly distributed on the observed
area.  This happens, for example, for data based on stars, which have
a higher density when one approaches the galactic equator.  However,
even for astronomical objects that are, in principle, uniformly
distributed on the sky (e.g., distant galaxies or quasars), we might
need to deal with a non-uniform distribution because of observational
effects (e.g., because of a non-constant sensitivity on the field of
the detector used, of dithering patterns, or of the presence on the
field of bright objects that interfere with the measurements).

The paper is organized as follow.  In Sect.~2 we carry out the various
generalizations in turn.  The properties of the average and the
two-point correlation function of the smoothed field are
discussed in Sect.~3.  Finally, we summarize the results of this paper
in Sect.~4.

\section{Main results}
\label{sec:main-results}

\subsection{Position dependent weight function}
\label{sec:posit-depend-weight}

Looking at Eq.~\eqref{eq:2}, we can note that the interpolation point
$\vec\theta$ does not directly enter the problem, but is basically only
used to ``label'' the interpolated point.  For our analysis, indeed,
it is convenient to fix that point to, say, $\vec\theta = \vec\theta_A$, and to
rewrite Eq.~\eqref{eq:2} as
\begin{equation}
  \label{eq:4}
  \tilde f_A \equiv \dfrac{\sum_{n=1}^N \hat f_n
  w_A(\vec\theta_n)}{\sum_{n=1}^N w_A(\vec\theta_n)} \; .
\end{equation}
where we called $f_A \equiv f(\vec\theta_A)$ and $w_A(\vec\theta') \equiv w(\vec\theta_A -
\vec\theta')$.  In this new notation, in the hypotheses used for Paper~I
(homogeneous Poisson process with uniform density $\rho$ for measurement
locations, infinite field), the expectation value $\bigl\langle \tilde
f_A\bigr\rangle$ can be evaluated exactly from the equations
\begin{align}
  \label{eq:5}
  Q_A(s) = {} & \rho \int_\Omega \bigl[ \e^{-s w_A(\vec\theta)} - 1
  \bigr] \, \diff^2 \theta \; , \\
  \label{eq:6}  
  Y_A(s) = {} & \exp \bigl[ Q_A(s) \bigr] \; , \\
  \label{eq:7}
  C_A(w_A) = {} & \frac{1}{1 - P_A} \int_0^\infty \e^{-w_A s} Y_A(s)
  \, \diff s \; , \\
  \label{eq:8}
  \bigl\langle \tilde f_A \bigr\rangle = {} & \rho \int_\Omega
  f(\vec\theta) w_A(\vec\theta) C_A\bigl( w_A(\vec\theta) \bigr) \,
  \diff^2 \theta \; ,
\end{align}
where $\Omega$ is the observation area (taken to be very large compared
with the typical scale of the weight function $w_A$) and $P_A$ is the
probability of having no point inside the support of $w_A$ (i.e.\ $P_A
= \exp(-\rho \pi_A)$, with $\pi_A$ the area of the support of $w_A$).

Equations (\ref{eq:4}--\ref{eq:8}) show explicitly that the location
$\vec\theta$ does not enter our problem.  As a result, looking again at
Eq.~\eqref{eq:2}, there is no need to take the weight function to be
of the form $w(\vec\theta - \vec\theta_n)$, but we can instead consider in the
same framework the more general form $w(\vec\theta, \vec\theta_n)$.  As we will
see below, the trivial generalization described in this subsection is
actually a fundamental step for more interesting results.

\subsection{Finite fields}
\label{sec:finite-fields}

We now focus on a slightly different generalization, namely the use of
finite fields.  We observe that having no data outside a given field
is totally equivalent to having data everywhere and using a vanishing
weight for locations outside the field.  In other words, even if our
observations are confined on a small part of the sky, we can always
imagine to have data in the whole sky, by generating arbitrary values
for the data locations and values, and then discard the arbitrary data
by using a vanishing weight for them.  In turn, from the form of the
integrands in Eqs.~\eqref{eq:5} and \eqref{eq:8}, we see that the
integrals actually need to be performed only on the domain of the
weight function (the integrands, indeed, vanish ot the points where
$w_A$ vanish).  As a result, Eqs.~(\ref{eq:5}--\ref{eq:8}) can still
be used, provided we interpret $\Omega$ as the observation field.

\subsection{Non-uniform density}
\label{sec:non-unif-dens}

We can now finally generalize Eqs.~(\ref{eq:5}--\ref{eq:8}) to
non-constant densities.  We first observe that the remaining spatial
variable $\vec\theta'$ that appears in the definition \eqref{eq:4} is
a dummy variable.  Indeed, $\vec\theta'$ is basically used to ``name''
locations, but does not really play any role on interpolation process.
For example, performing an arbitrary bijective (i.e., one-to-one)
mapping $\vec\theta \mapsto \vec\eta$ described by the function
$\vec\eta(\vec\theta)$, will not change the value of $\tilde f_A$,
provided that we use the ``mapped'' weight function
$w_A^{(\vec\theta)}(\vec\theta) = w_A^{(\vec\eta)}\bigl(
\vec\eta(\vec\theta) \bigr)$.  On the other hand, when performing the
transformation $\vec\theta \mapsto \vec\eta$, we are bound to change
the density distribution of objects.  More precisely, if the objects
are uniformly distributed on the plane $\vec\eta$ with density
$\rho^{(\vec\eta)}$, they will be distributed according to a
non-uniform density $\rho^{(\vec\theta)}(\vec\theta)$ on the
$\vec\theta$ plane.  The final density, indeed, is given by
\begin{equation}
  \label{eq:9}
  \rho^{(\vec\theta)}(\vec\theta) = \left\lvert \det \left(
  \frac{\partial \vec\eta(\vec\theta)}{\partial \vec\theta}
  \right)\right\rvert \, \rho^{(\vec\eta)} \; .
\end{equation}
This observation suggests a possible way to study non-uniform
densities.  Suppose that we intend to study the expectation value of
$\tilde f_A$ of Eq.~\eqref{eq:4} when the locations on the
$\vec\theta$ plane are distributed according to a non-uniform density
$\rho^{(\vec\theta)}(\vec\theta)$.  Then, we can look for a one-to-one
mapping $\vec\theta \mapsto \vec\eta$, such that the corresponding
density $\rho^{(\vec\eta)}$, evaluated according to Eq.~\eqref{eq:9},
is uniform; we also modify the weight function accordingly.  Now,
since in the $\vec\eta$ plane the locations are uniformly distributed,
we are in the position of studying the problem using the technique
developed in Paper~I.  Moreover, since the value of $\tilde f_A$ does
not depend on the coordinates $\vec\eta$ or $\vec\theta$ used, we can
directly use the result obtained.

The method described above clearly allows us to solve a much more
general problem but it has also two main problems.  From the
theoretical side, one has to show that it is possible to find a
one-to-one mapping that satisfies our needs (namely, that
$\rho^{(\vec\eta)}$ is uniform).  From the practical side, it might be
non-trivial to find the function $\vec\eta(\vec\theta)$; moreover, for
every point $\vec\theta_A$, one needs to transform the weight function
$w_A^{(\vec\theta)}(\vec\theta) = w(\vec\theta_A, \vec\theta)$ into a
weight function on the $\vec\eta$ plane (see above).  We now address
both problems, showing that our equations can be reformulated in a way
that naturally allows for non-uniform densities.

First, we explicitly show that, for every density distribution
$\rho^{(\vec\theta)}(\vec\theta)$ it is always possible to find a
one-to-one function $\vec\eta(\vec\theta)$ such that the corresponding
$\rho^{(\vec\eta)}$, evaluated from Eq.~\eqref{eq:9}, is constant.
Let us, in fact, consider the transformation
\begin{equation}
  \label{eq:10}
  \eta_i(\vec\theta) = 
  \begin{cases}
    \int_0^{\theta_1} \rho^{(\vec\theta)}(\theta'_1, \theta_2,
    \dotsc, \theta_M) \, \diff\theta'_1 & \text{if $i = 1$} \; , \\
    \theta_i & \text{otherwise} \; ,
  \end{cases}
\end{equation}
where $M$ is the number of dimensions of $\vec\theta$ (typically, in
Astrophysics, $M=2$).  This transformation satisfies the following
properties:
\begin{itemize}
\item It is continuous provided $\rho^{(\vec\theta)}$ is continuous.
  Situations where $\rho^{(\vec\theta)}$ is non-continuous will be
  considered at the end of this subsection.
\item It satisfies by construction Eq.~\eqref{eq:9}.  Indeed, the
  Jacobian matrix $\partial \vec\eta / \partial \vec\theta$ has the
  form
  \begin{equation}
    \label{eq:11}
    \frac{\partial \vec\eta}{\partial \vec\theta} = 
    \begin{pmatrix}
      \rho^{(\vec\theta)} & * & \hdotsfor{2} & * \\
      0 & 1 & 0 & \dots & 0 \\
      0 & 0 & 1 & \dots & 0 \\
      \vdots & \vdots & \vdots & \ddots & \vdots \\
      0 & 0 & 0 & \dots  & 1
    \end{pmatrix} \; ,
  \end{equation}
  where the stars ($*$) denotes uninteresting terms.
\item If $\rho^{(\vec\theta)}$ is strictly positive, the
  transformation is one-to-one.  Indeed, if $\rho^{(\vec\theta)} > 0$,
  then $\vec\eta(\vec\theta)$ is injective and also (clearly)
  subjective if restricted to its codomain.  In case
  $\rho^{(\vec\theta)}$ vanishes on some points, we can always use the
  argument of Sect.~\ref{sec:finite-fields}.  In other words, a
  vanishing density in a region implies that the probability of having
  points in this region vanishes as well; on the other hand, having no
  measurements on that region introduces a finite field described in
  Sect.~\ref{sec:finite-fields}.  Hence, we can assume that, inside
  $\Omega$, the density $\rho^{(\vec\theta)}$ is strictly positive.
\end{itemize}
This proofs the existence of a one-to-one function
$\vec\eta(\vec\theta)$ with the required properties (note that there
are many possible choices for $\vec\eta(\vec\theta)$, but they are all
equivalent for our purposes).

We now turn to the second problem, namely the practical difficulties
in applying the technique discussed in this section.  Suppose again
that we are interested in evaluating the expectation value of $\tilde
f_A$ of Eq.~\eqref{eq:4} with a non-uniform density
$\rho^{(\vec\theta)}(\vec\theta)$.  Then, we can use Eq.~\eqref{eq:10}
to convert the problem into the $\vec\eta$ plane, so that the
corresponding density is unity.  We can then finally apply
Eqs.~(\ref{eq:5}--\ref{eq:8}) on $\vec\eta$, using $\rho^{(\vec\eta)}
= 1$.  In particular, for Eq.~\eqref{eq:5} we have
\begin{align}
  \label{eq:12}
  Q_A(s) = {} & \rho^{(\vec\eta)} \int_{\vec\eta(\Omega)} \left[ \e^{-s
    w^{(\vec\eta)}_A(\vec\eta)} - 1 \right] \, \diff^2 \eta \notag\\
  {} = {} & \int_\Omega \left[ \e^{-s w^{(\vec\theta)}_A(\vec\theta)}
    - 1 \right] \left\lvert \det \left( \frac{\partial
        \vec\eta(\vec\theta)}{\partial \vec\theta} \right)\right\rvert
  \rho^{(\vec\eta)} \, \diff^2 \theta \notag\\
  {} = {} & \int_\Omega \left[ \e^{-s w^{(\vec\theta)}_A(\vec\theta)}
  - 1 \right] \rho^{(\vec\theta)}(\vec\theta) \, \diff^2 \theta \; .
\end{align}
Note that in the second line we have operated a change of variable
from $\vec\eta$ back to $\vec\theta$; note also that we have used the
fact that $w^{(\vec\theta)}(\vec\theta) = w^{(\vec\eta)}\bigl(
\vec\eta(\vec\theta) \bigr)$.  Hence, we can still use an equation
very close to Eq.~\eqref{eq:5}: we just need to integrate also over
the variable density $\rho^{(\vec\theta)}$.  Similarly, for
Eq.~\eqref{eq:8} we find
\begin{align}
  \label{eq:13}
  \bigl\langle \tilde f_A \bigr\rangle = {} & \rho^{(\vec\eta)} \int_{\vec\eta(\Omega)}
  f^{(\vec\eta)}_A(\vec\eta) w_A^{(\vec\eta)}(\vec\eta) C_A\Bigl(
  w_A^{(\vec\eta)}(\vec\eta) \Bigr) \, \diff^2 \eta \notag\\
  {} = {} & \int_\Omega f^{(\vec\theta)}_A(\vec\theta)
  w_A^{(\vec\theta)}(\vec\theta) C_A\Bigl(
  w_A^{(\vec\theta)}(\vec\theta) \Bigr) \left\lvert \det \left(
      \frac{\partial \vec\eta}{\partial \vec\theta}
    \right)\right\rvert \rho^{(\vec\eta)} \diff^2 \theta \notag\\
  {} = {} & \int_\Omega f^{(\vec\theta)}_A(\vec\theta)
  w_A^{(\vec\theta)}(\vec\theta) C_A\Bigl(
  w_A^{(\vec\theta)}(\vec\theta) \Bigr)
  \rho^{(\vec\theta)}(\vec\theta) \, \diff^2 \theta \; .
\end{align}
Equations~\eqref{eq:6} and \eqref{eq:7} remain unchanged.  Hence,
by simply using a transformation of variable back to $\vec\theta$, we
have been able to obtain a solution of the problem that does not
involve $\vec\eta$ any more.  This shows once more that the problem,
as expected, does not depend on the details of the choice of the
transformation $\vec\theta \mapsto \vec\eta$ (see comment at the end
of the previous paragraph).  Hence, in the following we will
drop the superscript $(\vec\theta)$ used in this section,
and hence we will always assume that all functions are evaluated in
the coordinate system defined by $\vec\theta$.  Finally, note that
the expressions in the last lines of Eq.~\eqref{eq:12} and
\eqref{eq:13} are still valid for vanishing densities; in other words,
we can either include the finite-field effects on the definition of
$\Omega$, or just put $\rho^{(\vec\theta)}(\vec\theta) = 0$ outside
the observation field.

Before closing this subsection, we consider the case of a
non-continuous density field $\rho^{(\vec\theta)}$.  We recall,
indeed, that with Eq.~\eqref{eq:10} we have been able to provide a
one-to-one, continuous transformation $\vec\eta(\vec\theta)$ only
under the hypothesis that $\rho^{(\vec\theta)}$ be continuous.
Although this hypothesis is not needed, it is non-trivial to exhibit a
mapping with the require properties in the general case of a
non-continuous density.  In reality, this problem is only apparent.
Note, in fact, that the density $\rho^{(\vec\theta)}$ enters
Eqs.~\eqref{eq:13} and \eqref{eq:15} only as a term inside an integral
and thus the continuity of this function does not play any role in our
problem.  For example, if we convolve a discontinuous density
$\rho^{(\vec\theta)}$ with a Gaussian,
\begin{equation}
  \label{eq:14}
  \rho^{(\vec\theta)}_\mathrm{s}(\vec\theta) = \int
  \frac{1}{2 \pi a^2} \exp\left( - \frac{\| \vec\theta'
  \|^2}{2 a^2} \right) \rho^{(\vec\theta)}(\vec\theta - \vec\theta')
  \, \diff^2 \theta' \; ,
\end{equation}
we obtain a smooth function
$\rho^{(\vec\theta)}_\mathrm{s}(\vec\theta)$.  This function, then,
can be used in Eqs.~\eqref{eq:13} and \eqref{eq:15} at the place of
the density.  Note that, since
$\rho^{(\vec\theta)}_\mathrm{s}(\vec\theta)$ is smooth for any $a >
0$, we can apply the transformation \eqref{eq:10} without any problem.
Finally, we take the limit $a \to 0^+$, so that the results of
the integrations \eqref{eq:13} and \eqref{eq:15} are not modified by
the use of $\rho^{(\vec\theta)}_\mathrm{s}$ instead of
$\rho^{(\vec\theta)}$.

\subsection{Average map}
\label{sec:final-solution}

\begin{figure}
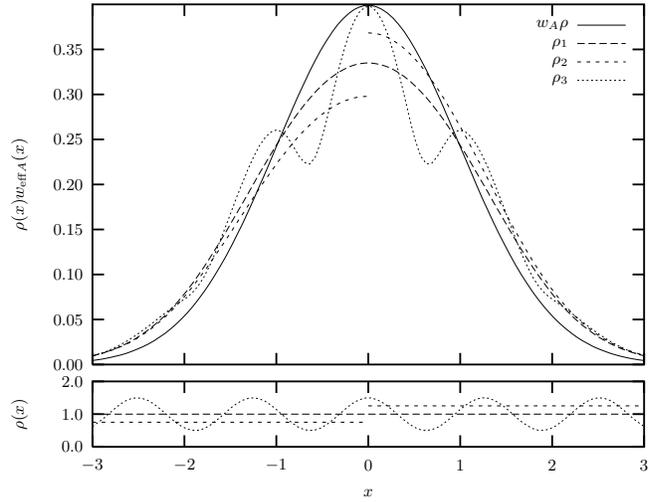

  \centering
  \resizebox{\hsize}{!}{\input fig1.tex}
  \caption{The effect of a non-constant density on the effective
    weight.  The plot shows, in the 1D case, the quantity $\rho(x)
    w_\mathrm{eff}(x)$ for three different densities, $\rho_1(x) = 1$,
    $\rho_2(x) = 1 + \mathrm{H}(x)/4$, and $\rho_3(x) = 1 - (\cos x) / 2$.
    In all cases the original weight function has been chosen to be
    such that the combination $w(x) \rho(x)$ is a Gaussian (see solid
    line plot).}
  \label{fig:1}
\end{figure}

\begin{figure}
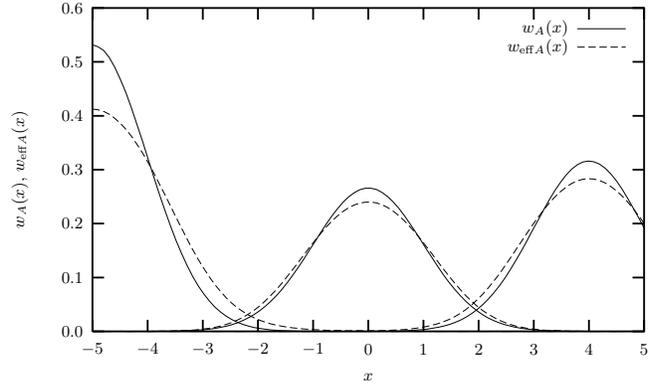

  \centering
  \resizebox{\hsize}{!}{\input fig2.tex}
  \caption{Effective weight function in presence of boundaries.  Three
    Gaussian weight function (shown in solid lines) centered on
    different part of the field $\Omega = [-5, 5]$ produce significantly
    different effective weights (dashed lines).  The weight function
    have been normalized according to Eq.~\eqref{eq:20}; $\rho(x) = 1.5$
    is constant for this plot.}
  \label{fig:2}
\end{figure}

We summarize here the results obtained in this section.  We have shown
that the expectation value of $\tilde f_A$ can be evaluated from the
set of equations
\begin{align}
  \label{eq:15}
  Q_A(s) = {} & \int_\Omega \bigl[ \e^{-s w_A(\vec\theta)} - 1 \bigr]
  \rho(\vec\theta) \, \diff^2 \theta \; , \\
  \label{eq:16}  
  Y_A(s) = {} & \exp \bigl[ Q_A(s) \bigr] \; , \\
  \label{eq:17}
  C_A(w_A) = {} & \frac{1}{1 - P_A} \int_0^\infty \e^{-w_A s} Y_A(s)
  \, \diff s \; , \\
  \label{eq:18}
  \bigl\langle \tilde f_A \bigr\rangle = {} & \int_\Omega
  f(\vec\theta) w_A(\vec\theta) C_A\bigl( w_A(\vec\theta) \bigr)
  \rho(\vec\theta) \, \diff^2 \theta \; ,
\end{align}
where the probability $P_A$ can be evaluated from
\begin{equation}
  \label{eq:19}
  P_A = \exp \biggl[ - \int_{\pi_{w_A}} \rho(\vec\theta) \, \diff^2
  \theta \biggr] \; .
\end{equation}
Appendix~A reports an alternative method to derive this set of
equations.  In the rest of the paper we will assume, without loss of
generality, that the weight function satisfies, for each point
$\vec\theta$, the (position dependent) normalization property
\begin{equation}
  \label{eq:20}
  \int_\Omega w(\vec\theta, \vec\theta') \rho(\vec\theta') \, \diff^2
  \theta' = \int_\Omega w_A(\vec\theta') \rho(\vec\theta') \, \diff^2
  \theta' = 1 \; .
\end{equation}
Indeed, since only \textit{relative\/} values of the weight function
are important in our problem, we can always suppose to deal with
weight functions normalized according to Eq.~\eqref{eq:20}.

Similarly to what was done in Paper~I, we call $w_{\mathrm{eff}A} =
w_A C_A(w_A)$ the \textit{effective weight function}, so that we can
write Eq.~\eqref{eq:18} as
\begin{equation}
  \label{eq:21}
  \bigl\langle \tilde f_A \bigr\rangle = \int_\Omega
  f(\vec\theta) w_{\mathrm{eff}A}(\vec\theta)
  \rho(\vec\theta) \, \diff^2 \theta \; .
\end{equation}
Note that, in contrast to Paper~I, in the definition of the effective
weight function we have not included the density $\rho(\vec\theta)$,
which thus must be explicitly added in the integration of
Eq.~\eqref{eq:21}.  This is convenient, because this way the effective
weight is a function of the \textit{value\/} of the
weight function and not (directly) of the position $\vec\theta$, and
because this way, as we will see in Sect.~\ref{sec:average-map},
the normalization property of the effective weight function is similar
to the normalization \eqref{eq:20} of the original weight.
The effects of a non-constant density and of a finite fields on the
effective weight are shown in Fig.~\ref{fig:1} and \ref{fig:2}.

\subsection{Final solution: two-point correlation function}
\label{sec:covariance}

We now turn to the generalization of the results of Paper~II
concerning the covariance of $\tilde f$, i.e.\ its two-point
correlation function.  Since the generalization procedure closely
follows the one used in Sect.~\ref{sec:main-results} for the average,
we skip here many details and mainly report the final result only.

We first recall that in Paper~II we have defined $\tilde
f_B$ and $w_B$ similarly with $\tilde f_A$ and $w_A$, with the only
difference that now these quantities are calculated with respect to a
different point $\vec\theta_B$.  We then have defined the two-point
correlation function of $\tilde f$ as
\begin{equation}
  \label{eq:22}
  \Cov(\tilde f; \theta_A, \theta_B) = \langle \tilde f_A \tilde f_B
  \rangle - \langle \tilde f_A \rangle \langle \tilde f_B \rangle \;
  ,
\end{equation}
and we have shown that this quantity is composed of two terms,
$\Cov(\tilde f; \theta_A, \theta_B) = T_\sigma + T_\mathrm{P}$, where $T_\sigma$ is the
noise due to measurement errors and $T_P = T_\mathrm{P1} +
T_\mathrm{P2} - T_\mathrm{P3}$ is the Poisson noise [split, in turn,
of three terms; see below Eqs.~(\ref{eq:27}--\ref{eq:29})].  

Using an argument similar to the one adopted in
Sect.~\ref{sec:main-results}, we can generalize the results of
Paper~II to the hypotheses discussed in the items of
Sect.~\ref{sec:introduction}.  We show here only the final results and
skip the proof, which is a trivial repetition of what was done above for
the average.
\begin{align}
  \label{eq:23}
  & Q(s_A, s_B) = \int_\Omega \bigl[ \e^{-s_A w_A(\vec\theta) - s_B
    w_B(\vec\theta)} - 1 \bigr] \rho(\vec\theta) \, \diff^2
  \theta \; , \\
  \label{eq:24}
  & Y(s_A, s_B) = \exp \bigl[ Q(s_A, s_B) \bigr] \; . \\
  \label{eq:25}
  & C(w_A, w_B) = \nu \int_0^\infty \! \diff s_A \int_0^\infty
  \! \diff s_B \, \e^{-s_A w_A - s_B w_B} Y(s_A, s_B) \; , \\
  \label{eq:26}
  & T_\sigma = \int_\Omega \diff^2 \theta \, \rho(\vec\theta)
  \sigma^2(\vec\theta) w_A(\vec\theta) w_B(\vec\theta) C \bigl(
  w_A(\vec\theta), w_B(\vec\theta) \bigr) \; , \\
  \label{eq:27}
  & T_\mathrm{P1} = \int_\Omega \diff^2 \theta \rho(\vec\theta) \bigl[
  f(\vec\theta) \bigr]^2 w_A(\vec\theta) w_B(\vec\theta) C\bigl(
  w_A(\vec\theta), w_B(\vec\theta) \bigr) \; , \\
  \label{eq:28}
  & T_\mathrm{P2} = \int_\Omega \! \diff^2 \theta_1 \,
  \rho(\vec\theta_1) \int_\Omega \! \diff^2 \theta_2 \,
  \rho(\vec\theta_2) f(\vec\theta_1) f(\vec\theta_2)
  w_A(\vec\theta_1) w_B(\vec\theta_2) \notag\\*
  & \phantom{T_\mathrm{P1} = {}} \times C\bigl( w_A(\vec\theta_1) +
  w_A(\vec\theta_2), w_B(\vec\theta_1) +
  w_B(\vec\theta_2) \bigr) \; , \\
  \label{eq:29}
  & T_\mathrm{P3} = \langle \tilde f_A \rangle \langle \tilde f_B
  \rangle \notag\\
  & \phantom{T_\mathrm{P3}} = \biggl[ \int_\Omega \diff^2 \theta_1 \,
  \rho(\vec\theta_1) f(\vec\theta_1) w_A(\vec\theta_1) C_A\bigl(
  w_A(\vec\theta_1) \bigr) \biggr] \notag\\*
  & \phantom{T_\mathrm{P3} = } \times \biggl[ \int_\Omega \diff^2
  \theta_2 \, \rho(\vec\theta_2) f(\vec\theta_2) w_B(\vec\theta_2)
  C_B\bigl( w_B(\vec\theta_2) \bigr) \biggl] \; .
\end{align}
We have called in Eq.~\eqref{eq:25} $\nu = 1/(1 - P_A - P_B + P_{AB})$,
where the probabilities $P_A$ and $P_B$ can be evaluated from
Eq.~\eqref{eq:19}, and $P_{AB}$ (the probability of having no points
inside both the supports of $w_A$ and $w_B$) is given by
\begin{equation}
  \label{eq:30}
  P_{AB} = \exp \biggl[ - \int_{\pi_{w_A} \cup \pi_{w_B}}
  \rho(\vec\theta) \, \diff^2 \theta \biggr] \; . 
\end{equation}

\section{Properties}
\label{sec:properties}

In this section we will consider some interesting properties of the
average (Sect.~\ref{sec:average-map}) and of the two-point correlation
function (Sect.~\ref{sec:two-point-corr}) of $\tilde f(\vec\theta)$.
Hence, here we basically generalize Sect.~5 of Paper~I and Sect.~6 of
Paper~II.

\subsection{Average map}
\label{sec:average-map}

\subsubsection{Normalization}
\label{sec:normalization}

By construction, for a constant field $f(\vec\theta) = 1$ the smoothed
function $\tilde f_A$ defined in Eq.~\eqref{eq:4} returns on average
$1$, a property related to the normalization of the effective weight
\citep[see][]{2002A&A...395..733L}.  Indeed, if $f(\vec\theta) = 1$, we
find
\begin{align}
  \label{eq:31}
  I & {} \equiv \langle \tilde f_A \rangle = \int_\Omega
  w_{\mathrm{eff}A}(\vec\theta) \rho(\vec\theta) \, \diff^2 \theta
  \notag\\
  & {} = \int_\Omega w_A(\vec\theta) C_A\bigl( w_A(\vec\theta) \bigr)
  \rho(\vec\theta) \, \diff^2 \theta \notag\\
  & {} = \frac{1}{1 - P_A} \int_0^\infty \diff s \, \e^{Q_A(s)} \int_\Omega
  \diff^2 \theta \, \rho(\vec\theta) w_A(\vec\theta) \e^{-s
  w_A(\vec\theta)} \; ,
\end{align}
where Eqs.~(\ref{eq:15}--\ref{eq:18}) have been used in the last step.
The last integral is just $-Q_A'(s)$, so that
\begin{align}
  \label{eq:32}
  I & {} = - \frac{1}{1- P_A} \int_0^\infty Q_A'(s) \e^{Q_A(s)} \,
  \diff s \notag\\
  & {} = \left. -\frac{1}{1 - P_A} \e^{Q_A(s)} \right|_0^\infty = 1 \; .
\end{align}
Hence, the effective weight has the same normalization property as the
weight function $w_A$ [see Eq.~\eqref{eq:20}].

\subsubsection{Scaling}
\label{sec:scaling}

Suppose we rescale the weight function $w_A(\vec\theta)$ into $k^2 w_A(k
\vec\theta)$, and at the same time the density $\rho(\vec\theta)$ into $k^2
\rho(\vec\theta)$; then we can verify using Eqs.~(\ref{eq:15}--\ref{eq:18})
that the effective weight is rescaled similarly to $w_A$, i.e.\ 
$w_{\mathrm{eff}A}(\vec\theta) \mapsto k^2 w_{\mathrm{eff}A}(k \vec\theta)$.

This scaling property suggests the following definition:
\begin{equation}
  \label{eq:33}
  \mathcal{N}_A \equiv \biggl[ (1 - P_A)
  \int_\Omega \bigl[ w_A(\vec\theta) \bigr]^2 \rho(\vec\theta) \,
  \diff^2 \theta \biggr]^{-1} \; .
\end{equation}
This quantity represents the number of ``relevant'' points used in the
average, i.e.\ the expected number of locations for which the weight
$w_A$ is significantly different from zero.  The $(1 - P_A)$ term in
Eq.~\eqref{eq:33} is introduced in order to compensate for
finite-fields effects [cf.\ the similar factor in Eq.~\eqref{eq:7}];
for example, for the top-hat function, this guarantees that
$\mathcal{N}_A = \rho \pi_{w_A} / \bigl[ 1 - \exp(-\rho \pi_{w_A}) \bigr] > 1$
for any density.  In any case, the above definition finds its main
justification from the properties that the quantity $\mathcal{N}_A$ so
defined enjoys (see below).  Following Paper~I, we call
$\mathcal{N}_A$ the \textit{weight number of objects\/}; similarly, we
define the \textit{effective weight number of objects\/}
$\mathcal{N}_{\mathrm{eff}A}$ using the effective weight
$w_{\mathrm{eff}A}$ instead of $w_A$ in Eq.~\eqref{eq:33}.  Note that
the related \textit{weight area\/} $\mathcal{A}$ and \textit{effective
  weight area\/} $\mathcal{A}_\mathrm{eff}$, also introduced in
Paper~I [see Eq.~(39) there], cannot be defined for the general case
of a non-uniform density.

\subsubsection{Behavior of $w_A C_A(w_A)$}
\label{sec:behavior-w-cw}

A study of $w_A C_A(w_A)$ can be carried out using the same technique
adopted in Paper~I.  Since $Y_A(s) > 0$ for every $s$, $C_A(w_A)$
decreases as $w_A$ increases.  Regarding $w_A C_A(w_A)$, from the
properties of Laplace transform (see, e.g., \citealp{Arfken}; see also
Appendix~D of Paper~II) we have
\begin{equation}
  \label{eq:34}
  w_{\mathrm{eff} A} = w_A C_A(w_A) = Y_A(0) + \Lp[Y'_A](w_A) \; ,
\end{equation}
where $\Lp[ \cdot ]$ indicates the Laplace transform.  Since $Y'_A(s) =
Q'_A(s) Y_A(s) < 0$, we find that $w_{\mathrm{eff}A}(w_A)$ increases
with $w_A$.  This implies that there must be a value $\bar w_A$ of the
weight function $w_A$ such that $C_A(w_A) > 1$ if $w_A < \bar w_A$,
and $C_A(w_A) < 1$ if $w_A > \bar w_A$.  Indeed, since $C_A$ is
monotonic, the equation $w_{\mathrm{eff}A}(w_A) = w_A$ can have at
most one solution; however, this equation must have at least one
solution because both $w_A$ and $w_{\mathrm{eff} A}$ satisfy the same
normalization property [cf.\ Eqs.~\eqref{eq:20} and \eqref{eq:32}].
The quantity
\begin{align}
  \label{eq:35}
  D \equiv \int_\Omega &\bigl[ w_A(\vec\theta) + w_{\mathrm{eff}
    A}(\vec\theta) - 2 \bar w_A \bigr] \notag\\
  & {} \times \bigl[ w_A(\vec\theta) - w_{\mathrm{eff}A}(\vec\theta)
  \bigr] \rho(\vec\theta) \, \diff^2 \theta  \geq 0 
\end{align}
is positive or null because the integrand is non-negative everywhere.
By expanding the integrand we find
\begin{align}
  \label{eq:36}
  0 \leq D = {} & \int_\Omega \bigl[ w_A(\vec\theta) \bigr]^2
  \rho(\vec\theta) \, \diff^2 \theta - \int_\Omega \bigl[
  w_{\mathrm{eff}A}(\vec\theta) \bigr]^2 \rho(\vec\theta) \, \diff^2
  \theta \notag\\
  & {} - 2 \bar w_A \int_\Omega \bigl[ w_A(\vec\theta) -
  w_{\mathrm{eff}A}(\vec\theta) \bigr] \rho(\vec\theta) \, \diff^2
  \theta \notag\\
  {} = {} & \frac{1}{1 - P_A} \left( \frac{1}{\mathcal{N}_A} -
    \frac{1}{\mathcal{N}_{\mathrm{eff}A}} \right) \; ,
\end{align}
where the normalization of $w_A$ and of $w_{\mathrm{eff}A}$ has been
used.  Hence, we find $\mathcal{N}_{\mathrm{eff}A} \geq \mathcal{N}_A$.

We now consider the limits of $w_{\mathrm{eff}A}(w_A)$ for small and
large values of $w_A$,
\begin{equation}
  \label{eq:37}
  \lim_{w_A \to \infty} w_A C_A(w_A) = \lim_{s \to
  0^+} \frac{Y(s)}{1 - P_A} = \frac{1}{1 - P_A} \; .
\end{equation}
Since $w_{\mathrm{eff}A}(w_A)$ is monotonic, $(1 - P_A)^{-1}$ is a
superior limit for the effective weight function.  This property, in
turn, can be used inside the definition of
$\mathcal{N}_{\mathrm{eff}A}$ to obtain
\begin{align}
  \label{eq:38}
  \mathcal{N}_{\mathrm{eff}A}^{-1} = {} & (1 - P_A) \int_\Omega \bigl[
  w_A(\vec\theta) \bigr]^2 \rho(\vec\theta) \, \diff^2 \theta \notag\\
  {} < & \int_\Omega w_A(\vec\theta) \rho(\vec\theta) \, \diff^2 \theta = 1
  \; .
\end{align}
In other words, the effective number of objects will always exceed
unity, independently of the weight function used.

Regarding the other limit we have
\begin{equation}
  \label{eq:39}
  \lim_{w_A \to 0^+} w_A C_A(w_A) = \lim_{s \to \infty}
  \frac{1}{1 - P_A} Y_A(s) = \frac{P_A}{1 - P_A} \; .
\end{equation}
Hence, since $w_{\mathrm{eff}A}(w_A)$ vanishes if $w_A = 0$, the
effective weight has a discontinuity at $0$ if $P_A \neq 0$.

\subsubsection{Limit of high and low densities}
\label{sec:limit-high-low}

At high densities ($\rho \to \infty$) only values of $Q_A(s)$ close to $s = 0$
are important, because for large $s$, $Y_A(s)$ vanishes.  Hence, we
expand $Q_A(s)$ by writing
\begin{equation}
  \label{eq:40}
  Q_A(s) = \sum_{k=1}^\infty \frac{(-1)^k s^k S_{Ak}}{k!} \; ,
\end{equation}
where $S_{Ak}$ is the $k$th moment of $w_A$:
\begin{equation}
  \label{eq:41}
  S_{Ak} \equiv \int_\Omega \bigl[ w_A(\vec\theta) \bigr]^k
  \rho(\vec\theta) \, \diff^2 \theta \; .
\end{equation}
The normalization \eqref{eq:20} implies $S_{A1} = 1$, and so to first
order $Y_A(s) \simeq \e^{-s}$.  We have then
\begin{equation}
  \label{eq:42}
  C_A(w_A) \simeq \frac{1}{1 - P_A} \int_0^\infty \e^{-s w_A} \e^{-s} \,
  \diff s = \frac{1}{1 - P_A} \frac{1}{1 + w_A} \; .
\end{equation}

In the limit of low densities ($\rho \to 0^+$), instead, $Y_A(s) \to 1$ and
\begin{equation}
  \label{eq:43}
  C_A(w_A) \simeq \frac{1}{1 - P_A} \frac{1}{w_A} \; .
\end{equation}
Expanding Eq.~\eqref{eq:19} to first order in $\rho$ we find, for $w_A
> 0$,
\begin{equation}
  \label{eq:44}
  w_{A\mathrm{eff}} = w_A C_A(w_A) \simeq \biggl[ \int_{\pi_{w_A}}
  \rho(\vec\theta) \, \diff^2 \theta \biggr]^{-1} \; .
\end{equation}
Hence, the effective weight converges to a top-hat function normalized
to unity.

\subsubsection{Moments expansion}
\label{sec:moments-expansion}

At large densities $\rho$, we can expand $C_A(w_A)$ in terms of the
moments of $w_A$ defined in Eq.~\eqref{eq:41}.  Calculations are
basically identical to the one provided in Paper~I [see Eq.~(66) of
that paper], with only minor corrections due to the different
definition of $C_A$.  Hence, we skip the derivation and report here
only the final result (up to the fifth term):
\begin{align}
  \label{eq:45}
  (1 - P_A) C_A(w_A) \simeq {} & \frac{1}{1 + w_A} + \frac{S_{A2}}{(1 + w_A)^3}
  - \frac{S_{A3}}{(1 + w_A)^4} \notag\\*
  & {} + \frac{S_{A4} + 3 S_{A2}^2}{(1 + w_A)^5} \; .
\end{align}

\subsection{Two-point correlation function}
\label{sec:two-point-corr}

\begin{figure}
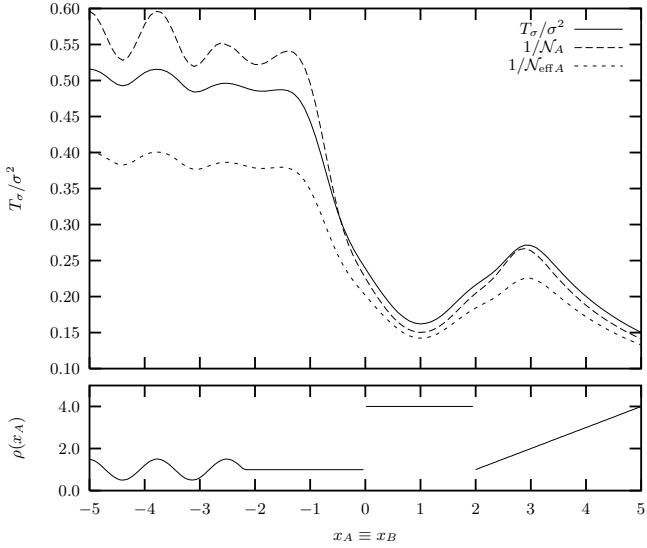

  \centering
  \resizebox{\hsize}{!}{\input fig3.tex}
  \caption{The variance $T_\sigma$ evaluated at different positions $x_A \equiv
    x_B$. For this plot, we used a Gaussian weight function with
    variance $1/2$; the measurement error $\sigma$ was kept constant on the
    field.  Note that, in agreement with Eq.~\eqref{eq:51}, the
    quantity $T_\sigma/\sigma^2$ is always larger than
    $1/\mathcal{N}_\mathrm{eff}$; curiously, the quantity
    $1/\mathcal{N}$ gives a good first-order approximation for $T_\sigma$
    even in the complex situation shown here.}
  \label{fig:3}
\end{figure}

The generalization of the properties of the covariance terms $T_\sigma$ and
$T_\mathrm{P}$ is, in most cases, trivial and closely follows the
generalization carried out in the Sect.~\ref{sec:average-map}.  Hence,
here we will skip much of the proofs and just outline the main
results.

\subsubsection{Normalization}
\label{sec:normalization-1}

It can be shown that the Poisson noise satisfies a simple
normalization property: if $f(\vec\theta) = 1$ is constant, then
$T_\mathrm{P1} + T_\mathrm{P2} = T_\mathrm{P3} = 1$, and thus the
Poisson noise vanishes.  A proof of this property can be carried out
either with the technique described in Paper~II, or, more easily,
using the following argument, taken from \citet{2002A&A...395..733L}.
If $f(\vec\theta) = 1$ is constant on the field, we will on average measure
$\langle \hat f_n \rangle = 1$ for each point.  Let us now assume for a moment
that we do not have any measurement error, so that $\tilde f_n = 1$
for every $n$.  Then, we will always measure $\tilde f(\vec\theta) = 1$,
and thus $\bigl\langle f(\vec\theta_A) \bigr\rangle = \bigl\langle f(\vec\theta_B) \bigr\rangle = \bigl\langle
f(\vec\theta_A) f(\vec\theta_B) \bigr\rangle = 1$.  In this case, thus, we find
$T_\mathrm{P1} + T_\mathrm{P2} = T_\mathrm{P3} = 1$.  The situation is
actually the same even if the measurements are affected by errors:
These, in fact, appear only in the evaluation of $T_\sigma$, and thus the
Poisson noise is left unaffected.

\subsubsection{Small and large separations}
\label{sec:large-separations}

In the limit $\vec\theta_A \equiv \vec\theta_B$, the expressions for $T_\sigma$ and
$T_\mathrm{P}$ take a particularly simple form.  Indeed, since $w_A \equiv
w_B$, to evaluate $T_\sigma$, $T_\mathrm{P1}$, and $T_\mathrm{P2}$ we just
need $C(w_A, w_A)$; this quantity, in turn, can be easily shown to be
$C(w_A, w_A) = -C'_A(w_A)$, where $C_A(w_A)$ is given by
Eq.~\eqref{eq:17}.

If instead $\lvert \vec\theta_A - \vec\theta_B \rvert$ is large compared to the
scale lengths of the weight functions $w_A$ and $w_B$, then
\begin{align}
  \label{eq:46}
  Q(s_A, s_B) \simeq {} & Q_A(s_A) + Q_B(s_B) \; , \\
  \label{eq:47}
  Y(s_A, s_B) \simeq {} & Y_A(s_A) Y_B(s_B) \; , \\
  \label{eq:48}
  C(w_A, w_B) \simeq {} & C_A(w_A) C_B(w_B) \; .
\end{align}
The following argument shows that in general $C(w_A, w_B) \geq C_A(w_A)
C_B(w_B)$.  First, observe that, since $P_{AB} \geq P_A P_B$, one has $\nu
\geq (1 - P_A)^{-1} (1 - P_B)^{-1}$.  Moreover, it can be shown that
$Q(s_A, s_B) \geq Q_A(s_A) + Q_B(s_B)$: indeed
\begin{align}
  \label{eq:49}
  & Q(s_A, s_B) - Q_A(s_A) - Q_B(s_B) \notag\\
  & \quad {} = \int_\Omega \bigl[ \e^{-s_A w_A(\vec\theta)} - 1 \bigr]
  \bigl[ \e^{-s_B w_B(\vec\theta)} - 1 \bigr] \rho(\vec\theta) \,
  \diff^2 \theta \geq 0 \; .
\end{align}
Finally then
\begin{align}
  \label{eq:50}
  & C(w_A, w_B) - C_A(w_A) C_B(w_B) \notag\\*
  & \quad {} \geq \frac{1}{1 - P_A} \frac{1}{1 - P_B} \int_0^\infty
  \diff s_A \int_0^\infty \diff s_B \, \e^{-s_A w_A - s_B w_B} \notag\\*
  & \qquad \times \left[ \e^{Q(s_A, s_B)} - \e^{Q_A(s_A) + Q_B(s_B)}
  \right] \geq 0 \; ,
\end{align}
where the last inequality is a consequence of the monotonicity of the
exponential function and of Eq.~\eqref{eq:49}.

\subsubsection{Behaviour of $T_\sigma$}
\label{sec:behaviour-t_sigma}

The normalization of the Poisson noise terms derived in
Sect.~\ref{sec:normalization-1} can be used to derive an upper limit
for $T_\sigma$.  Indeed, from the expression of $T_\sigma$ one sees that this
quantity is very similar to $T_\mathrm{P1}$, provided we replace
$f(\vec\theta)$ with $\sigma(\vec\theta)$.  On the other hand, from
Sect.~\ref{sec:normalization-1} we know that, if $f(\vec\theta) = 1$, then
$T_\mathrm{P1} < 1$, because in this case $T_\mathrm{P1} +
T_\mathrm{P2} = 1$ and $T_\mathrm{P2}$ is positive.  Hence we find
$T_\sigma < \sigma^2_\mathrm{max}$, where $\sigma^2_\mathrm{max}$ is an upper limit
for $\sigma^2(\vec\theta)$.

A lower limit for $T_\sigma$ can be obtained from the inequality
\eqref{eq:50}:
\begin{align}
  \label{eq:51}
  T_\sigma {} & \geq \int_\Omega \sigma^2(\vec\theta) w_A(\vec\theta)
  w_B(\vec\theta) C_A\bigl( w_A(\vec\theta) \bigr) C_B\bigl(
  w_B(\vec\theta) \bigr) \rho(\vec\theta) \, \diff^2
  \theta \notag\\
  & {} = \int_\Omega \sigma^2(\vec\theta)
  w_{\mathrm{eff}A}(\vec\theta) w_{\mathrm{eff}B}(\vec\theta)
  \rho(\vec\theta) \, \diff^2 \theta \; .
\end{align}
Note that if $\theta_A \equiv \theta_B$, we have $w_{\mathrm{eff}A}(\vec\theta) \equiv
w_{\mathrm{eff}B}(\vec\theta)$ and the last integral in Eq.~\eqref{eq:51}
is closely related to $\mathcal{N}_{\mathrm{eff}A} \equiv
\mathcal{N}_{\mathrm{eff}B}$.  The two limits on $T_\sigma$ discussed in
this section are exemplified in Fig.~\ref{fig:3}.

\subsubsection{Limit of high and low densities}
\label{sec:limit-high-low-1}

At high densities we can expand $Q(s_A, s_B)$ as done in
Sect.~\ref{sec:limit-high-low} for $Q_A$:
\begin{align}
  \label{eq:52}
  Q(s_A, s_B) & {} = \int \biggl[ \sum_{i,j} \frac{1}{i! j!} \bigl[
  -s_A w_A(\vec\theta) \bigr]^i \bigl[ -s_B w_B(\vec\theta) \bigr]^j -
  1 \biggr] \notag\\*
  & \quad {} \times \rho(\vec\theta) \, \diff^2 \theta \simeq - s_A -
  s_B \; ,
\end{align}
where in the last step we have retained only the first terms of the
sum, and used the normalization of $w_A$ and $w_B$ [cf.\
Eq.~\eqref{eq:20}].  Hence, in this limit 
\begin{equation}
  \label{eq:53}
  C(w_A, w_B) \simeq \nu \frac{1}{1 + w_A} \frac{1}{1 + w_B} \; .
\end{equation}
Note that, because of the normalization \eqref{eq:20}, both weight
functions $w_A$ and $w_B$ behave like $\rho^{-1}$ at high densities,
and thus $C(w_A, w_B) \simeq 1$ to first order.  We then find
\begin{equation}
  \label{eq:54}
  T_\sigma \simeq \int_\Omega \sigma^2(\vec\theta) w_A(\vec\theta)
  w_B(\vec\theta) \rho(\vec\theta) \, \diff^2 \theta \; .
\end{equation}
This expression should be compared with Eq.~\eqref{eq:36}.

If instead $\rho \to 0^+$, then $Q(s_A, s_B) \to 0^-$, $Y(s_A, s_B) \to 1$ and
thus $C(w_A, w_B) \to \nu / (w_A w_B)$.  Note that here we are assuming
$w_A(\vec\theta) \neq 0$ and $w_B(\vec\theta) \neq 0$.  In the same limit, we have
[see Eqs.~\eqref{eq:19} and \eqref{eq:30}]
\begin{align}
  \label{eq:55}
  \nu^{-1} & {} \simeq \biggl( \int_{\pi_{w_A}} + \int_{\pi_{w_B}} -
  \int_{\pi_{w_A} \cup \pi_{w_B}} \biggr) \rho(\vec\theta) \, \diff^2
  \theta \notag\\
  & {} = \int_{\pi_{w_A} \cap \pi_{w_B}} \rho(\vec\theta) \, \diff^2
  \theta \; .
\end{align}
Hence, we finally find
\begin{equation}
  \label{eq:56}
  T_\sigma \simeq \biggl[ \int_{\pi_{w_A} \cap \pi_{w_B}} \mskip-30mu
  \sigma^2(\vec\theta) \rho(\vec\theta) \, \diff^2 \theta \biggr] \biggm/
  \biggl[ \int_{\pi_{w_A} \cap \pi_{w_B}} \mskip-30mu \rho(\vec\theta)
  \, \diff^2
  \theta \biggr] \; .
\end{equation}
This result is consistent with the upper limit for $T_\sigma$ obtained in
Sect.~\ref{sec:behaviour-t_sigma}.

\section{Conclusion}
\label{sec:conclusion}

In this paper we have studied the statistical properties of a
smoothing technique widely used in Astronomy and in other physical
sciences.  In particular, we have provided simple analytical
expressions to calculate the average and the two-point correlation
function of the smoothed field $\tilde f(\vec\theta)$ defined in
Eq.~\eqref{eq:1}.  The results generalize what was already obtained in
Paper~I and II to the case where observations are carried out in a
finite field, with a non-uniform spatial density for the measurements,
and with non-uniform measurement errors $\sigma^2(\vec\theta)$.  These
generalizations together greatly widen the range of applicability of
our results in the astronomical context.  Finally, we have shown
several interesting properties of the average map and of the two-point
correlation function, and we have considered the behavior of these
quantities in some relevant limiting cases.

\acknowledgements{This work was partially supported by a grant from
  the Deutsche Forschungsgemeinschaft, and the TMR Network
  ``Gravitational Lensing: New constraints on Cosmology and the
  Distribution of Dark Matter.''}

\appendix

\section{Alternative derivation}
\label{sec:altern-deriv}

In this appendix, we derive the same results obtained in
Sect.~\ref{sec:main-results} using a more direct method.  Although not
necessary, this alternative derivation is helpful in order to fully
understand the whole problem and also clarifies some of the
peculiarities of the equations derived in Paper~I (cf., in particular,
the case of vanishing weights).

The derivation will follow quite closely the one adopted in Paper~I,
with the needed modifications due to the finite-field and the
non-constant density.  The only significant exception will be the use
of a different strategy in performing the so-called ``continuous
limit'' (because of the finite field, we cannot perform the limit $N
\to \infty$, but we must rather take $N$ as a random
variable).  Note that throughout this appendix we will drop everywhere
the index $A$, so that, e.g., $w_A(\vec\theta)$ will be written just
as $w(\vec\theta)$.  This simplification should not create
ambiguities, since anyway here we are concerned only with the value of
$\tilde f$ at $\vec\theta_A$.

Let us consider a field $\Omega$ and locations randomly distributed on
this field with density $\rho(\vec\theta)$.  Let us assume, for
simplicity, that $w(\vec\theta)$ is strictly positive in $\Omega$;
if this is not the case, we can always redefine $\Omega$ to include
only points inside the support of $w$ (cf.\ discussion in
Sect.~\ref{sec:finite-fields}).  The expected average number of
locations in $\Omega$ is given by
\begin{equation}
  \label{eq:57}
  \bar N = \int_\Omega \rho(\vec\theta) \, \diff^2 \theta \; .
\end{equation}
The actual number of points will be a random variable following a
Poisson distribution with average $\bar N$.  In reality, since we are
accepting only cases where there is at least one point inside the
support of $w$ (we could not define $\tilde f$ otherwise), $N$ will
follow the probability
\begin{equation}
  \label{eq:58}
  p_N(N) = \frac{\e^{-\bar N}}{1 - \e^{-\bar N}} \frac{\bar N^N}{N!} \;
  ,
\end{equation}
where $N$ is assumed to be a positive integer.  Note in particular
that the normalization factor takes into account the ``missing'' $N=0$
probability.

Since the locations are distributed inside $\Omega$ according to the
density $\rho(\vec\theta)$, a single location follows the probability
distribution $\rho(\vec\theta) / \bar N$; note that the factor $1 /
\bar N$ is needed here in order to satisfy the normalization of
probabilities (the integral on $\Omega$ must be unity).  Hence, the
probability of having exactly $N$ locations inside $\Omega$ at the
positions $\{ \vec\theta_n \}$ (with $n \in \{ 1, \dotsc, N\}$) is
given by
\begin{equation}
  \label{eq:59}
  P\bigl( \{ \vec\theta_n \} \bigr) = p_N(N) \prod_{n=1}^N
  \frac{\rho(\vec\theta_n)}{\bar N} = \frac{1}{\e^{\bar N} - 1}
  \frac{1}{N!} \prod_{n=1}^N \rho(\vec\theta_n) \; .
\end{equation}
We can now use this probability distribution to evaluate the
expectation value of $\tilde f$ defined in Eq.~\eqref{eq:4}:
\begin{align}
  \label{eq:60}
  \langle \tilde f \rangle = {} & \sum_{N=1}^\infty \int_\Omega
  \diff^2 \theta_1 \dotsi \int_\Omega \diff^2 \theta_N P\bigl( \{
  \vec\theta_n
  \} \bigr) \notag\\*
  & \phantom{\sum_{N=1}^\infty \int_\Omega} {} \times
  \frac{\sum_{n=1}^N f(\vec\theta_n) w(\vec\theta_n) }{\sum_{n=1}^N
    w(\vec\theta_n)} \notag\\
  {} = {} & \frac{1}{\e^{\bar N} - 1} \sum_{N=1}^\infty \frac{1}{N!}
  \int_\Omega \diff^2 \theta_1 \, \rho(\vec\theta) \dotsi \int_\Omega
  \diff^2 \theta_N \, \rho(\vec\theta_N) \notag\\*
  & \phantom{\frac{1}{\e^{\bar N} - 1} \sum_{N=1}^\infty \frac{1}{N!}
    \int_\Omega} \times \frac{N f(\vec\theta_1)
    w(\vec\theta_1)}{\sum_{n=1}^N w(\vec\theta_n)} \; .
\end{align}
Similarly to Paper~I, we now define, for each $N$, the random variables
\begin{equation}
  \label{eq:61}
  y_N \equiv \sum_{n=2}^N w(\vec\theta_n) \; .
\end{equation}
In order to evaluate the associated probability distribution, we
consider the probability distribution for values of the weight $w$
(i.e., we use Marcov's method; see, e.g.,
\citealp{1943RvMP...15....1C, 1987PhRvL..59.2814D}).  The probability
density for a location to have a weight $w$, in particular, is given
by
\begin{equation}
  \label{eq:62}
  p_w(w) = \frac{1}{\bar N} \int_\Omega \diff^2 \theta
  \rho(\vec\theta) \delta \bigl( w - w(\vec\theta) \bigr) \; .
\end{equation}
This allows us to evaluate the probability distribution for $y_N$ as
\begin{align}
  \label{eq:63}
  p_{y_N} = {} & \frac{1}{\bar N^{N - 1}} \int_\Omega \diff^2
  \theta_2 \, \rho(\vec\theta_2) \dotsi \int_\Omega \diff^2
  \theta_N \, \rho(\vec\theta_N) \notag\\*
  & \phantom{\frac{1}{\bar N^{N - 1}} \int_\Omega} \times \delta\bigl(
  y_N - w(\vec\theta_2) - \dotsb - w(\vec\theta_N) \bigr) \notag\\
  {} = {} & \int_0^\infty \diff w_2 \, p_w(w_2) \dotsi
  \int_0^\infty \diff w_N \, p_w(w_N) \notag\\*
  & \phantom{\int_0^\infty} \times \delta(y_N - w_2 - \dotsb -
  w_N) \; .
\end{align}
Using this probability, we can rewrite Eq.~\eqref{eq:60} as
\begin{align}
  \label{eq:64}
  \langle \tilde f \rangle = {} & \frac{1}{\e^{\bar N} - 1}
  \sum_{N=1}^\infty \frac{\bar N^{N-1}}{N!} \int_\Omega \diff^2
  \theta_1 \, \rho(\vec\theta_1) w(\vec\theta_1)
  f(\vec\theta_1) \notag\\*
  & {} \times \int_0^\infty \frac{N p_{y_N}(y_N) \, \diff
    y_N}{w(\vec\theta_1) + y_N} \; .
\end{align}
The form of this expression justifies the definition
\begin{equation}
  \label{eq:65}
  C(w) \equiv \frac{1}{\e^{\bar N} - 1} \sum_{N=1}^\infty \frac{\bar
  N^{N-1}}{N!} \int_0^\infty \frac{N p_{y_N}(y_N) \, \diff
    y_N}{w + y_N} \; ,
\end{equation}
so that we have
\begin{equation}
  \label{eq:66}
  \langle \tilde f \rangle = \int_\Omega \diff^2 \theta_1 \,
  \rho(\vec\theta_1) w(\vec\theta_1) f(\vec\theta_1) C\bigl(
  w(\vec\theta_1) \bigr) \; .
\end{equation}
In order to further simplify the definition of $C$, we use a technique
similar to the one adopted in Paper~I.  Namely, we define $W$, the
Laplace transform of $p_w$
\begin{align}
  \label{eq:67}
  W(s) \equiv {} & \Lp[p_w](s) = \int_0^\infty \e^{-s w}
  p_w(w_A) \, \diff w \notag\\
  {} = {} & \frac{1}{\bar N} \int_\Omega \e^{-s w(\vec\theta)}
  \rho(\vec\theta) \, \diff^2 \theta \; .
\end{align}
We similarly define, for each $N$, $Y_N$, the Laplace transform of
$p_{y_N}$, as
\begin{align}
  \label{eq:68}
  Y_N(s) = {} & \Lp[p_{y_N}](s) = \int_0^\infty
  p_{y_N}(y_N) \e^{-s y_N} \, \diff y_N \notag\\
  {} = {} & \bigl[ W(s) \bigr]^{N-1} \; .
\end{align}
We now use the following property of Laplace transforms: if $f$ is any
function, and $x_0$ any positive real number, we have [see Eqs.~(14),
(20), and (25) of Paper~I for a proof]
\begin{equation}
  \label{eq:69}
  \Lp\bigl[ \Lp[f] \bigr](x_0) = \int_0^\infty \frac{f(x)}{x_0 + x} \,
  \diff x \; .
\end{equation}
Using this in Eq.~\eqref{eq:65} we find
\begin{align}
  \label{eq:70}
  C(w) = {} & \frac{1}{\e^{\bar N} - 1} \sum_{N=1}^\infty
  \frac{\bar N^{N - 1}}{N!} N \Lp[y_N](w) \notag\\
  {} = {} & \frac{1}{\e^{\bar N} - 1} \Lp \biggl[ \sum_{N=1}^\infty
  \frac{\bar N^{N - 1}}{(N-1)!} W^{N-1} \biggr](w) \notag\\
  {} = {} & \frac{1}{\e^{\bar N} - 1} \Lp \biggl[ \sum_{\nu=0}^\infty
  \frac{\bar N^\nu}{\nu!} W^\nu \biggr](w) \notag\\
  {} = {} & \frac{1}{\e^{\bar N} - 1}
  \Lp\bigl[ \e^{\bar N W} \bigr](w) \notag\\
  {} = {} & \frac{1}{1 - \e^{-\bar N}} \Lp\bigl[ \e^{\bar N W - \bar
    N} \bigr](w) \; .
\end{align}
Finally, we define $Q$ as
\begin{equation}
  \label{eq:71}
  Q(s) \equiv \bar N W(s) - \bar N = \int_\Omega \bigl[ \e^{-s
  w(\vec\theta)} - 1 \bigr] \rho(\vec\theta) \, \diff^2 \theta \; ,
\end{equation}
so that we finally have
\begin{equation}
  \label{eq:72}
  C(w) = \frac{1}{1 - \e^{-\bar N}} \Lp \bigl[\e^{Q(s)}
  \bigr](w) \; .
\end{equation}
This completes our proof.

\bibliographystyle{aa}
\bibliography{../lens-refs}

\end{document}